\documentclass[aps,prl,reprint,superscriptaddress]{revtex4-1}

\newcommand{\GRad}{\Gamma_{\mathrm{Rad}}}
\newcommand{\GMix}{\Gamma_{\mathrm{Mix}}}
\newcommand{\GTTwo}{\Gamma_{T_2}}
\newcommand{\GAdd}{\Gamma_\mathrm{Add}}

\newcommand{\tRad}{\tau_\mathrm{Rad}}

\newcommand{\GAone}{\Gamma_{A_1}}
\newcommand{\GAtwo}{\Gamma_{A_2}}
\newcommand{\GEonetwo}{\Gamma_{E_{1,2}}}
\newcommand{\As}{|^1A_1\rangle}
\newcommand{\At}{\,^3A_2}
\newcommand{\Es}{|^1E_{1,2}\rangle}
\newcommand{\Et}{\,^3E}
\newcommand{\Eonetwo}{|E_{1,2}\rangle}
\newcommand{\Ex}{|E_x\rangle}
\newcommand{\Ey}{|E_y\rangle}

\newcommand{\Aone}{|A_1\rangle}
\newcommand{\Atwo}{|A_2\rangle}

\newcommand{\msz}{m_s=0}
\newcommand{\mso}{|m_s|=1}

\usepackage{graphicx,xr,natbib,hyperref,amsmath,amsthm,amssymb,wasysym,epstopdf,cleveref}
\usepackage[svgnames]{xcolor}
\hypersetup{hidelinks,colorlinks=true,allcolors=DarkBlue,pdfpagemode=UseNone}

\begin{document}

\title{Phonon-Induced Population Dynamics and Intersystem Crossing\\in Nitrogen-Vacancy Centers}

\author{M. L. Goldman}
\email[]{mgoldman@physics.harvard.edu}
\affiliation{Department of Physics, Harvard University, Cambridge, Massachusetts 02138, USA}

\author{A. Sipahigil}
\affiliation{Department of Physics, Harvard University, Cambridge, Massachusetts 02138, USA}

\author{M. W. Doherty}
\affiliation{Laser Physics Centre, Research School of Physics and Engineering, Australian National University, Australian Capital Territory 0200, Australia}

\author{N. Y. Yao}
\affiliation{Department of Physics, Harvard University, Cambridge, Massachusetts 02138, USA}

\author{S. D. Bennett}
\affiliation{Department of Physics, Harvard University, Cambridge, Massachusetts 02138, USA}

\author{M. Markham}
\affiliation{Element Six Ltd, Kings Ride Park, Ascot SL5 8BP, UK}

\author{D. J. Twitchen}
\affiliation{Element Six Ltd, Kings Ride Park, Ascot SL5 8BP, UK}

\author{N. B. Manson}
\affiliation{Laser Physics Centre, Research School of Physics and Engineering, Australian National University, Australian Capital Territory 0200, Australia}

\author{A. Kubanek}
\affiliation{Department of Physics, Harvard University, Cambridge, Massachusetts 02138, USA}

\author{M. D. Lukin}
\affiliation{Department of Physics, Harvard University, Cambridge, Massachusetts 02138, USA}



\begin{abstract}
We report direct measurement of population dynamics in the excited state manifold of a nitrogen-vacancy (NV) center in diamond.  We quantify the phonon-induced mixing rate and demonstrate that it can be completely suppressed at low temperatures.  Further, we measure the intersystem crossing (ISC) rate for different excited states and develop a theoretical model that unifies the phonon-induced mixing and ISC mechanisms.  We find that our model is in excellent agreement with experiment and that it can be used to predict unknown elements of the NV center's electronic structure.  We discuss the model's implications for enhancing the NV center's performance as a room-temperature sensor.
\end{abstract}


\maketitle

The nitrogen-vacancy (NV) center in diamond has emerged as a versatile atomlike system, finding diverse applications in metrology and quantum information science at both ambient and cryogenic temperatures.  At room temperature, the NV center has broad appeal as a sensor---e.g. for nanoscale, biocompatible thermometry \cite{Kucsko2013}, magnetometry \cite{Grinolds2013}, pressure sensing \cite{Doherty2014}, and electric field sensing \cite{Dolde2011}.  NV centers also have the potential to serve as quantum registers featuring high-fidelity quantum gates \cite{VanderSar2012} and second-long coherence times \cite{Maurer2012a}.  All of these applications depend critically on a spin-dependent nonradiative transition into a metastable state, the so-called intersystem crossing (ISC), which enables nonresonant optical initialization and readout of the electronic spin state.  Despite theoretical \cite{Manson2006} and experimental \cite{Manson2006,Batalov2008,Robledo2011,Toyli2012} efforts, a detailed understanding of the microscopic ISC mechanism has remained an open question.  Such an understanding may enable efforts to enhance the NV center's optical initialization and readout fidelities, or to identify or engineer similar mechanisms in other solid-state defects.

\begin{figure}
\begin{center}
\includegraphics[width=\columnwidth]{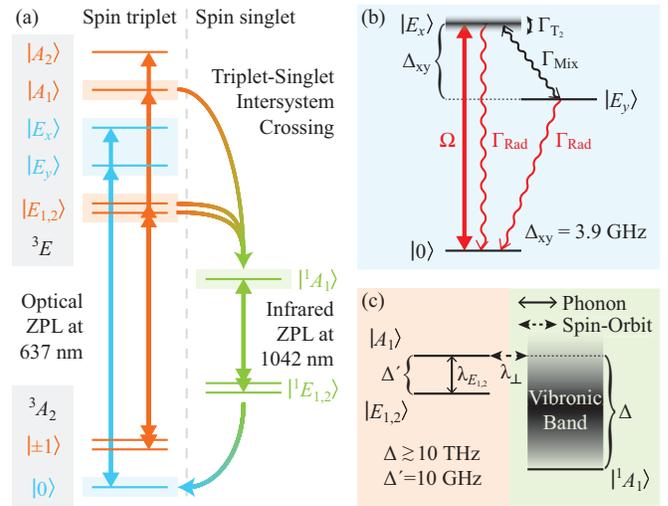}
\caption{The level structure of the NV center. (a) A schematic illustration of the NV center's level structure. The intersystem crossing (ISC) process is responsible for shelving into $\Es$ via the short-lived $\As$ and for pumping into $|0\rangle$. (b) The three-level system of spin-triplet, $m_s=0$ states used to model phonon-induced mixing and dephasing within the $\Et$ manifold. (c) The states, Hamiltonian matrix elements ($\lambda_{E_{1,2}}$, $\lambda_\perp$), and energy scales ($\Delta$, $\Delta^\prime$) involved in the triplet-singlet ISC.}
\label{fig.Level Structure}
\end{center}
\end{figure}

In this Letter, we use resonant optical manipulation of an NV center at cryogenic temperatures to probe the NV center's interaction with phonons in the diamond lattice.  The NV center has a spin-triplet, orbital-singlet ground state ($\At$) that is coupled optically to a spin-triplet, orbital-doublet excited state ($\Et$), as shown in \crefformat{figure}{Fig.~#2#1{(a)}#3}\cref{fig.Level Structure}.  We investigate the ISC from the $\Et$ manifold to the intermediate spin-singlet states ($\As$, $\Es$) and phonon-mediated population transfer within the $\Et$ manifold.  We first measure phonon-induced population transfer between $\Ex$ and $\Ey$, using the effectively closed three-level system shown in \crefformat{figure}{Fig.~#2#1{(b)}#3}\cref{fig.Level Structure}, to characterize phononic coupling to the orbital electronic state.  We build on previous such measurements \cite{Fu2009,Kaiser2009}, in which population transfer was not shown to be completely suppressed, by adopting techniques that enable highly coherent excitation of the NV center's optical transitions \cite{Robledo2010,Bernien2013}.

We next measure the fluorescence lifetimes of $\Aone$, $\Atwo$, and $\Eonetwo$ and find that the ISC rates from these states differ sharply.  Based on these experimental observations, we develop a model of the ISC mechanism that combines spin-orbit coupling, phonon-induced electronic state transitions, and phonon-mediated lattice relaxation.  Using our measured phonon-induced mixing rate as an input, we find excellent quantitative agreement between our model, our experimental results, and previous observations \cite{Manson2006,Batalov2008,Robledo2011,Toyli2012}.

In our experiment, we use a 1.0 mm diameter solid-immersion lens (SIL) that is fabricated from bulk electronic grade CVD diamond and cut along the (100) crystal plane \cite{Siyushev2010}.  The SIL is mounted in a continuous flow helium cryostat that allows us to vary the temperature from 4.8 K to room temperature.  We use a laser at 532 nm for nonresonant initialization of the NV center's charge and spin states, and two tuneable external-cavity diode lasers gated by electro-optical amplitude modulators to apply independent resonant pulses at 637 nm.  Using photoluminescence excitation (PLE) spectroscopy, we resolve five of the six dipole-allowed $\At\rightarrow\Et$ transitions; at zero applied magnetic field and low strain, $|E_1\rangle$ and $|E_2\rangle$ are too close in energy for their transitions from $|\pm1\rangle$ to be resolved.

To measure the mixing rate between $\Ex$ and $\Ey$, we measure both the decay rate out of one of the states and the rate of population transfer between the two states.  First, we measure the decay rate out of $\Ex$ by measuring the timscale $\tau_\mathrm{Rabi}$ on which optical Rabi oscillations between $|0\rangle$ and $\Ex$ decohere.  We apply resonant 60 ns pulses and record the arrival times of the resulting phonon sideband (PSB) photons, as the rate of spontaneous emission into the PSB is instantaneously proportional to the population in $\Ex$ \cite{SuppInfo,Robledo2010}.

\crefformat{figure}{Fig.~#2#1{(b)}#3}
\begin{figure}
\begin{center}
\includegraphics[width=\columnwidth]{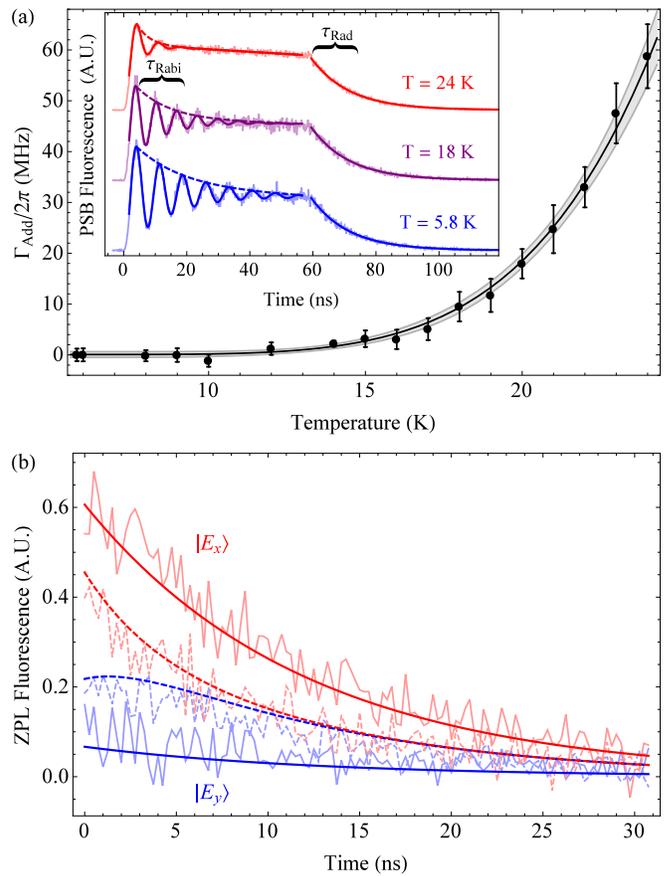}
\caption{Phonon-induced mixing between the $\Ex$ and $\Ey$ electronic orbital states.
(a) The measured $\GAdd=\GMix+\GTTwo$ as a function of temperature, with a fit to $\GAdd\propto T^5$ \cite{SuppInfo}.  The shaded region is the $95\%$ confidence interval and the inset shows Rabi oscillations on the $|0\rangle\rightarrow\Ex$ transition measured at three temperatures, offset for clarity.
(b) Background-subtracted fluorescence of $x$ (red) or $y$ (blue) polarization collected after resonant excitation to $\Ex$. Data were taken at $T=5.0$ K (solid lines) and $T=20$ K (dashed lines). The fits are simulations to the three-level system depicted in \cref{fig.Level Structure} \cite{SuppInfo}.}
\label{fig.Ex-Ey Mixing}
\end{center}
\end{figure}

For each dataset, we fit $\tau_\mathrm{Rabi}$ from the oscillation decay and extract $\GRad=1/\tRad$ from the pulse's falling edge, where $\tRad$ is the radiative lifetime of $\Ex$.  From these two values, we extract
\begin{equation}
\label{eq.Additional Rabi decoherence}
\GAdd=\GMix+\GTTwo=2\left(1/\tau_\mathrm{Rabi}-\frac{3}{4}\GRad\right),
\end{equation}
the additional decoherence rate of the Rabi oscillations due to processes other than optical decay to $|0\rangle$ \cite{Cohen-Tannoudji1992,SuppInfo}.  $\GMix$ and $\GTTwo$ are the phonon-induced mixing and dephasing rates, respectively.  Typical Rabi oscillations and the derived values of $\GAdd$ are shown in \crefformat{figure}{Fig.~#2#1{(a)}#3}\cref{fig.Ex-Ey Mixing}.  The $T^5$ scaling of $\GAdd$ indicates that the additional Rabi decoherence is due primarily to mixing between $\Ex$ and $\Ey$ mediated by two $E$-symmetric phonons \cite{Goldman2014a,Fu2009}.  We infer $\GAdd/2\pi=-0.34\pm1.87$ MHz at 5.8 K ($95\%$ confidence interval), indicating that phonon-induced mixing is frozen out at low temperature.  There also exists a one-phonon emission process whose contribution to the mixing rate scales as $\Delta_{xy}^2T$ \cite{Goldman2014a}.  This contribution is negligible in our experiment because of the small density of states for phonons of frequency $\Delta_{xy}=3.9$ GHz \footnote{Although the one-phonon process contributes negligibly in this experiment, it may cause significant mixing in NV centers, such as those formed by nitrogen implantation or placed inside nanofabricated structures, where damage to the local crystalline structure may induce a large strain splitting.  We note that this linear temperature scaling may be evident in the result of a previous measurement of phonon-induced mixing, which used NV centers with strain splittings of 8 to 81 GHz \cite{Fu2009}.}.

We also measure population transfer between $\Ex$ and $\Ey$ directly by measuring the depolarization of the emitted zero-phonon line (ZPL) fluorescence.  ZPL photons emitted by decay from the $\Ex$ and $\Ey$ states have orthogonal linear polarizations (labeled $x$ and $y$).  At 5 K and 20 K, we resonantly excite the NV center to $\Ex$ and collect fluorescence of $x$ and $y$ polarizations, as shown in \crefformat{figure}{Fig.~#2#1{(b)}#3}\cref{fig.Ex-Ey Mixing}.  At both temperatures, the emission is $x$-polarized for small delays, indicating initial decay primarily from $\Ex$.  The emission remains $x$-polarized at 5 K, whereas we observe that emission becomes depolarized at 20 K.  Since emission polarization is directly related to excited state population, this is a direct observation of population transfer between $\Ex$ and $\Ey$. We compare the observed population transfer to simulations of rate equations based on the three-level system depicted in \crefformat{figure}{Fig.~#2#1{(b)}#3}\cref{fig.Level Structure} \cite{SuppInfo}. Using the values of $\GAdd$ given by the fit in \crefformat{figure}{Fig.~#2#1{(a)}#3}\cref{fig.Ex-Ey Mixing}, and using our polarization selectivity and the starting time of the mixing/radiative decay dynamics relative to the excitation pulse as fit parameters, we find good agreement between the observed and simulated fluorescence depolarization \cite{SuppInfo}.

\begin{figure}
\begin{center}
\includegraphics[width=\columnwidth]{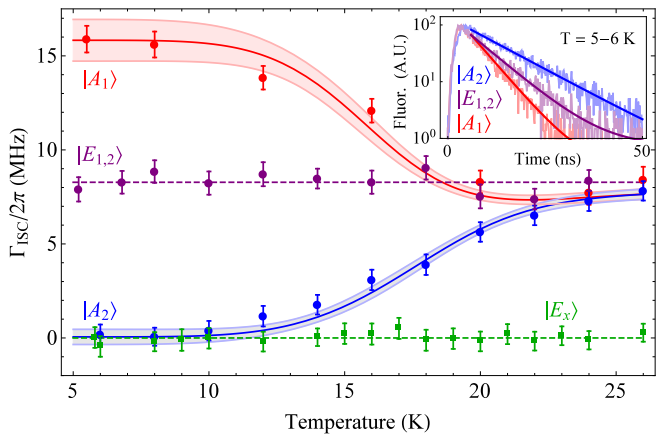}
\caption{ISC rates from the $\Et$ triplet excited states.  The inset shows the measured PSB fluorescence collected after excitation to $\Aone$, $\Atwo$, and $\Eonetwo$ measured at $T\sim5$ K, normalized to a common peak height and fit to an exponential decay curve.  The blue and red bands are fits with $95\%$ confidence intervals to the phonon-induced mixing model described in the text, and the purple and green lines are placed at the mean values of the corresponding data sets.}
\label{fig.Lifetimes vs Temperature}
\end{center}
\end{figure}

We now turn from the $\Et$ states with $m_s=0$ ($\Ex$ and $\Ey$) to the states with $|m_s|=1$ ($\Aone$, $\Atwo$, and $|E_{1,2}\rangle$).  Although the radiative decay rate $\GRad$ is the same for all $\Et\rightarrow\At$ transitions \cite{Batalov2008,Robledo2011}, one expects from symmetry arguments that $\Aone$, $\Atwo$, and $\Eonetwo$ should exhibit different ISC rates into the spin-singlet states \cite{Doherty2013a}.  We can therefore probe population dynamics among these states by exciting the NV center into one state and measuring the decay time of the resulting PSB fluorescence as a function of temperature.  A representative measurement is shown in the inset to Fig. \ref{fig.Lifetimes vs Temperature}.  From the fluorescence decay time $\tau_i$ measured after excitation into the $i$th state, we can calculate the associated ISC rate
\begin{equation}
\label{eq.Gamma_ISC}
\Gamma_i=1/\tau_i-\GRad.
\end{equation}
Because the ISC rates from $\Ex$ and $\Ey$ are negligible ($\Gamma_{E_x}/2\pi\leq0.62\pm0.21$ MHz \cite{SuppInfo}) compared to the ISC rates from $\Aone$, $\Atwo$, and $\Eonetwo$, we set $\GRad/2\pi=1/2\pi\bar{\tau}_{E_x} = 13.2\pm0.5$ MHz, where $\bar{\tau}_{E_x}$ is the average lifetime of $\Ex$ extracted from our Rabi decoherence data.  The derived values of $\Gamma_i$ are shown in Fig. \ref{fig.Lifetimes vs Temperature}.

\begin{figure}
\begin{center}
\includegraphics[width=\columnwidth]{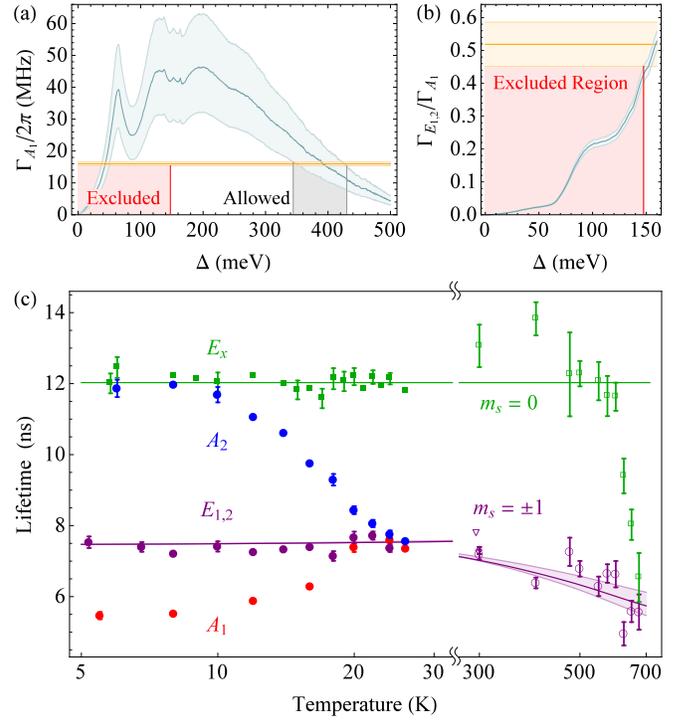}
\caption{The results of the ISC model. (a) The range (blue shading) of predicted values of $\GAone$ corresponding to the range of possible values of $\lambda_\perp$.  The orange line is the measured value of $\GAone$, the black region indicates our extracted value of $\Delta$ (the $\Aone-\As$ energy spacing), and the red region indicates the values of $\Delta$ that are excluded by the measured $\GEonetwo/\GAone$ ratio.
(b) The predicted (blue) and measured (orange) $\GEonetwo/\GAone$ ratio.  Because we assume no acoustic phonon cutoff, the blue plot represents an upper bound on the ratio for a given $\Delta$.  We can therefore exclude the values of $\Delta$ indicated by the red region, which is reproduced in (a).
(c) The lifetimes of the $m_s=0$ (green) and $|m_s|=1$ (purple) states predicted for a range of temperatures.  The data from 295 K to 700 K are taken from Refs. \cite{Toyli2012} ($\Box$ and $\ocircle$), \cite{Robledo2011} ($\diamond$), and \cite{Batalov2008} ($\triangledown$).}
\label{fig.Model Results}
\end{center}
\end{figure}

We observe that the $\Aone$, $\Atwo$, and $\Eonetwo$ ISC rates are significantly different at low temperatures, but converge around $T\gtrsim22$ K.  The same two-phonon process that redistributes population among $\Ex$ and $\Ey$ also does so among $\Aone$ and $\Atwo$.  As a result, the observed temperature-dependent ISC rates ($\tilde{\Gamma}_{A_1}$ and $\tilde{\Gamma}_{A_2}$) converge to an average of the two unmixed states' rates ($\GAone$ and $\GAtwo$) as the temperature increases.  We fit $\tilde{\Gamma}_{A_1}$ and $\tilde{\Gamma}_{A_2}$, assuming $\GAtwo=0$ and a temperature-dependent $\Aone-\Atwo$ mixing rate equal to the measured $\Ex-\Ey$ rate \cite{SuppInfo}.  We find excellent agreement using only $\GAone$ as a free parameter, confirming that the same phonon-induced mixing process is evident in both Figs. \ref{fig.Ex-Ey Mixing} and \ref{fig.Lifetimes vs Temperature}.  The state lifetimes we observe at $T\geq22$ K are consistent with those of the $m_s=0$ and $|m_s|=1$ states at room temperature [see \crefformat{figure}{Fig.~#2#1{(c)}#3}\cref{fig.Model Results}], indicating that we have measured the onset of the orbital averaging mechanism that enables the $\Et$ manifold to be treated as an effective spin-triplet, orbital-singlet system at room temperature \cite{Fuchs2008,Rogers2009}.

We now present a theoretical analysis of the ISC mechanism, which is treated in greater detail in Ref. \cite{Goldman2014a}.  In the NV center, an axial spin-orbit (SO) interaction ($\propto \lambda_{||}l_z s_z$) is primarily responsible for the fine structure of the $\Et$ manifold while a transverse SO interaction [$\propto \lambda_\perp \left(l_x s_x + l_y s_y\right)$] couples states of different spin multiplicities \cite{Doherty2011,Maze2011}.  The ISC occurs in two steps [see \crefformat{figure}{Fig.~#2#1{(c)}#3}\cref{fig.Level Structure}]: (1) a SO-mediated transition from a state in the $\Et$ manifold to a resonant excited vibrational level of $\As$, and (2) relaxation of the excited vibrational level to the ground (or thermally occupied) vibrational level of $\As$.  Because the latter occurs on the picosecond timescale \cite{Huxter2013}, the overall ISC rate is defined by the initial SO-mediated transition.

According to Fermi's golden rule, the SO-mediated transition requires both SO coupling and overlap between the initial vibrational level of $\Et$ and the excited vibrational level of $\As$. Because $\As$ and the $\At$ states consist of the same single-particle electronic orbitals, the two states exhibit similar charge density distributions and therefore similar vibrational potentials \cite{Kehayias2013}.  The vibrational overlap between the $\Et$ states and $\As$ is then well approximated by that observed in the PSB of the $\Et\rightarrow\At$ optical emission spectrum.  Selection rules imply that only $\Aone$ is SO-coupled with $\As$.  Thus, SO coupling can mediate a first-order transition from $\Aone$ to a resonant excited vibrational level of $\As$, whereas $\Eonetwo$ must undergo a second-order transition involving electron-phonon coupling with $\Aone$ [see \crefformat{figure}{Fig.~#2#1{(c)}#3}\cref{fig.Level Structure}]. In principle, $\Eonetwo$ may undergo a first-order transition to a highly excited vibrational level of $\Es$, but we expect the rate of this transition to be negligible because the $\As-\Es$ energy spacing (1190 meV \cite{Acosta2010a}) is large compared to the extent of the PSB ($\sim500$ meV \cite{Davies1976,Kehayias2013}).

The ISC rate from $\Aone$ is
\begin{align}
\label{eq.A1 ISC}
\Gamma_{A_1} &= 4\pi \hbar \, \lambda_\perp^2 \sum\limits_n |\langle \chi_{0} | \chi^\prime_{\nu_n} \rangle|^2 \: \delta\hspace{-1 pt}\left(\nu_n - \Delta\right) \nonumber \\
&= 4\pi \hbar \, \lambda_\perp^2 \: F\hspace{-1.5 pt}\left(\Delta\right),
\end{align}
where $\lambda_\perp$ is the transverse spin-orbit coupling rate, $\Delta$ is the energy spacing between $\Aone$ and $\As$, and $\delta$ is the Dirac delta function. $|\chi_0\rangle$ is the ground vibrational level of $\Aone$, and $|\chi^\prime_{\nu_n}\rangle$ are the vibrational levels of $\As$ with energies $\nu_n$ above that of $\As$.  We define the vibrational overlap function $F\hspace{-1.5 pt}\left(\Delta\right) = \overline{|\langle \chi_0 | \chi^\prime_{\Delta} \rangle|^2} \; \rho\hspace{-1.5 pt}\left(\Delta\right)$, where the average is over all vibrational levels with energy $\Delta$ and $\rho\hspace{-1.5 pt}\left(\Delta\right)$ is the associated density of states.

At low temperature, the ISC rate from $\Eonetwo$ is
\begin{align}
\label{eq.E12 ISC}
\Gamma_{E_{1,2}} \hspace{-1 pt} &= 4\pi \hbar^3 \lambda_\perp^2 \hspace{-2 pt} \sum\limits_{n,p,k} \hspace{-1 pt} \frac{\lambda_{p,k}^2}{\omega_k^2}
 \: |\langle \chi_0 | \chi^\prime_{\nu_n} \rangle|^2 \:
 \delta\hspace{-1 pt}\left(\nu_n + \omega_k - \Delta\right) \nonumber \\
&= \frac{2}{\pi} \, \hbar \, \eta \: \Gamma_{A_1} \hspace{-13 pt} \int\limits_0^{\min\left(\Delta,\Omega\right)} \hspace{-12 pt} \omega \: \frac{F\hspace{-1.5 pt}\left(\Delta-\omega\right)}{F\hspace{-1.5 pt}\left(\Delta\right)}
\: \mathrm{d}\omega,
\end{align}
where $\lambda_{p,k}$ is the phononic coupling rate for a phonon of polarization $p$ and wavevector $k$, and
\begin{equation}
J\hspace{-1.5 pt}\left(\omega\right)=\frac{\pi\hbar}{2}\sum\limits_k \lambda_{E_{1,2},k}^2\:\delta\left(\omega-\omega_k\right)=\eta \:\omega^3,
\end{equation}
is the phonon spectral density in the acoustic limit \cite{Fu2009} for the polarization that couples $\Eonetwo$ with $\Aone$.  We assume a cutoff energy $\Omega$ for acoustic phonons.

Because the SO interaction is $A_1$-symmetric \cite{Lenef1996}, and can therefore only couple states of like symmetry, $\Atwo$ is not SO-coupled to either $\As$ or $\Es$ \cite{Doherty2011,Maze2011}.  Similar symmetry considerations forbid single-phonon coupling between $\Atwo$ and $\Aone$, so neither the first- nor second-order processes described above can induce ISC decay from $\Atwo$.  The lowest-order allowed mechanism would be a third-order process involving two phonons and one SO interaction, but we neither expect nor observe an appreciable ISC transition rate due to such a high-order process.

In \crefformat{figure}{Fig.~#2#1{(a)}#3}\cref{fig.Model Results}, we plot the prediction of Eq. \ref{eq.A1 ISC} as a function of $\Delta$ and the value of $\GAone/2\pi=16.0\pm0.6$ MHz extracted from the fits shown in Fig. \ref{fig.Lifetimes vs Temperature}.  The broad range of predicted values of $\GAone$ arises from the currently imprecise knowledge of the transverse SO coupling rate $\lambda_\perp$ \footnote{The axial SO coupling rate \unexpanded{$\lambda_{||}=5.33\pm0.03$} GHz is known \cite{Bassett2014}, but the precise value of \unexpanded{$\lambda_\perp/\lambda_{||}\approx1$} remains an open question.  As explained in Ref. \cite{Goldman2014a}, we have selected \unexpanded{$\lambda_\perp/\lambda_{||}=1.2\pm0.2$} as a reasonable confidence interval.}.  The vibrational overlap function is extracted from a previous measurement of the $\Et\rightarrow\At$ emission PSB \cite{Kehayias2013}.  The intersection of the measured and predicted values of $\GAone$ confines $\Delta$ to two regions: around 43 meV and from 344 to 430 meV.

We next evaluate $\GEonetwo/\GAone$, which depends only on electron-phonon coupling parameters and the vibronic overlap function, not on the SO coupling rate.  We extract $\eta=2\pi\times\left(44.0\pm2.4\right)\,\mathrm{MHz}\:\mathrm{meV}^{-3}$, which parameterizes the electron-phonon coupling strength, from the $\Ex-\Ey$ mixing data shown in \crefformat{figure}{Fig.~#2#1{(a)}#3}\cref{fig.Ex-Ey Mixing} \cite{SuppInfo}.  In \crefformat{figure}{Fig.~#2#1{(b)}#3}\cref{fig.Model Results}, we plot the ratios predicted by Eqs. \ref{eq.A1 ISC}-\ref{eq.E12 ISC}
\footnote{The value of \unexpanded{$\GEonetwo$} calculated in \cref{fig.Model Results} includes a correction to Eq. \ref{eq.E12 ISC} due to a second-order ISC process that uses \unexpanded{$\Es$} as intermediate states instead of \unexpanded{$\Aone$}.  This correction, which is described explicitly in Ref. \cite{Goldman2014a}, lowers \unexpanded{$\GEonetwo/\GAone$} by \unexpanded{$21\%$} at \unexpanded{$\Delta=148$} meV.}
and extracted from Fig. \ref{fig.Lifetimes vs Temperature}.  We assume no acoustic cutoff energy ($\Omega\rightarrow\infty$) in order to maximize the range of acoustic phonon modes that contribute to $\GEonetwo$, making the predicted ratio an upper bound.  Even so, we find that the predicted and measured ratios are inconsistent for $\Delta<148$ meV, which uniquely confines $\Delta$ to the region from 344 to 430 meV \footnote{In this region, we find that the predicted and measured ratios match for cutoff energies of 74 to 93 meV.  These values, which are below the Debye energy (194 meV \cite{Bosak2005}) and close to the energies of the quasi-local phonon modes that dominate the \unexpanded{$\Et$} (64 meV \cite{Kehayias2013,Davies1974}) and \unexpanded{$\As$} (71 meV \cite{Davies1976}) PSBs, are physically reasonable.}.

We scale our model up to higher temperatures \cite{Goldman2014a} and find that, for this range of $\Delta$, its predictions are consistent with published lifetimes of the $\msz$ and $\mso$ $\Et$ states at temperatures between 295 K and 600 K, as shown in \crefformat{figure}{Fig.~#2#1{(c)}#3}\cref{fig.Model Results}.  An additional decay mechanism, which is not captured by our model of ISC decay, significantly shortens the lifetime of the $\msz$ states above 600 K.  This decay mechanism is discussed further in Ref. \cite{Goldman2014a}.



We have elucidated, both experimentally and theoretically, the roles that electron-phonon interactions play in NV center dynamics.  Further exploration of either of the phonon roles addressed in this Letter may yield intriguing applications.  Resonant electron-phonon coupling in the $\Et$ manifold could be used to optically cool a high-Q diamond resonator \cite{Ovartchaiyapong2012,Tao2014} close to the vibrational ground state \cite{Kepesidis2013a}.  Such efforts would complement the growing interest in using electron-phonon coupling in the $\At$ states to manipulate the electron spin \cite{MacQuarrie2013,Ovartchaiyapong2014a} or to generate spin-squeezed states of NV ensembles \cite{Bennett2013}.  Further, our understanding of the ISC mechanism may enable efforts to engineer the ISC rate by, for example, applying a large static strain to shift the energy spacings between the spin-triplet and -singlet states \cite{Doherty2014}.  Such an advance would provide an across-the-board enhancement to the spin initialization and readout techniques on which room-temperature NV center applications depend.  Finally, our experimentally validated ISC model has confined the unknown energy of the $\Et$ and $\As$ states to a region that can be explored in future optical spectroscopy.

\begin{acknowledgments}
The authors would like to thank J. Maze, A. Gali, E. Togan, A. Akimov, D. Sukachev, Q. Unterreithmeier, A. Zibrov, and  A. Palyi for stimulating discussions and experimental contributions. This work was supported by NSF, CUA, the DARPA QUEST program, AFOSR MURI, ARC, Element Six, and the Packard Foundation. A. K. acknowledges support from the Alexander von Humboldt Foundation.
\end{acknowledgments}

\crefformat{figure}{Fig.~#2#1{(b)}#3}

%

\end{document}


\newcommand{\figSize}{0.5}
\newcommand{\GRad}{\Gamma_{\mathrm{Rad}}}
\newcommand{\GMix}{\Gamma_{\mathrm{Mix}}}
\newcommand{\GTTwo}{\Gamma_{T_2}}
\newcommand{\GAdd}{\Gamma_\mathrm{Add}}
\newcommand{\GISC}{\Gamma_\mathrm{ISC}}
\newcommand{\GAone}{\Gamma_{A_1}}
\newcommand{\GAtwo}{\Gamma_{A_2}}
\newcommand{\GEonetwo}{\Gamma_{E_{1,2}}}
\newcommand{\tAdd}{\tau_\mathrm{Add}}
\newcommand{\tRad}{\tau_\mathrm{Rad}}
\newcommand{\tRabi}{\tau_\mathrm{Rabi}}
\newcommand{\As}{|^1A_1\rangle}
\newcommand{\At}{\,^3A_2}
\newcommand{\Es}{|^1E_{1,2}\rangle}
\newcommand{\Et}{\,^3E}
\newcommand{\Eonetwo}{|E_{1,2}\rangle}
\newcommand{\Ex}{|E_x\rangle}
\newcommand{\Ey}{|E_y\rangle}
\newcommand{\Aone}{|A_1\rangle}
\newcommand{\Atwo}{|A_2\rangle}
\newcommand{\dAoa}{\delta_{A_1^a}}
\newcommand{\dAob}{\delta_{A_1^b}}
\newcommand{\dEoa}{\delta_{E_1^a}}
\newcommand{\dEob}{\delta_{E_1^b}}
\newcommand{\dEta}{\delta_{E_2^a}}
\newcommand{\dEtb}{\delta_{E_2^b}}
\newcommand{\dEota}{\delta_{E_{1,2}^a}}
\newcommand{\pAoa}{\hat{P}_{A_1^a}}
\newcommand{\pAob}{\hat{P}_{A_1^b}}
\newcommand{\pEoa}{\hat{P}_{E_1^a}}
\newcommand{\pEob}{\hat{P}_{E_1^b}}
\newcommand{\pEta}{\hat{P}_{E_2^a}}
\newcommand{\pEtb}{\hat{P}_{E_2^b}}
\newcommand{\msz}{m_s=0}
\newcommand{\mspmo}{|m_s|=1}

\newcommand{\GRadx}{\Gamma_\mathrm{Rad}^{(x)}}
\newcommand{\GRady}{\Gamma_\mathrm{Rad}^{(y)}}
\newcommand{\Gxy}{\Gamma_\mathrm{Mix}^{(x)}}
\newcommand{\Gyx}{\Gamma_\mathrm{Mix}^{(y)}}
\newcommand{\GISCx}{\Gamma_{\mathrm{ISC},E_x}}
\newcommand{\GISCone}{\Gamma_{\mathrm{ISC},\pm1}}
\newcommand{\GISCzero}{\Gamma_{\mathrm{ISC},0}}

\renewcommand{\thefigure}{S\arabic{figure}}
\renewcommand{\theequation}{S\arabic{equation}}

\title{Supplemental Material: Phonon-Induced Population Dynamics and Intersystem Crossing in Nitrogen-Vacancy Centers}

\author{M. L. Goldman}
\email[]{mgoldman@physics.harvard.edu}
\affiliation{Department of Physics, Harvard University, Cambridge, Massachusetts 02138, USA}

\author{A. Sipahigil}
\affiliation{Department of Physics, Harvard University, Cambridge, Massachusetts 02138, USA}

\author{M. W. Doherty}
\affiliation{Laser Physics Centre, Research School of Physics and Engineering, Australian National University, Australian Capital Territory 0200, Australia}

\author{N. Y. Yao}
\affiliation{Department of Physics, Harvard University, Cambridge, Massachusetts 02138, USA}

\author{S. D. Bennett}
\affiliation{Department of Physics, Harvard University, Cambridge, Massachusetts 02138, USA}

\author{M. Markham}
\affiliation{Element Six Ltd, Kings Ride Park, Ascot SL5 8BP, UK}

\author{D. J. Twitchen}
\affiliation{Element Six Ltd, Kings Ride Park, Ascot SL5 8BP, UK}

\author{N. B. Manson}
\affiliation{Laser Physics Centre, Research School of Physics and Engineering, Australian National University, Australian Capital Territory 0200, Australia}

\author{A. Kubanek}
\affiliation{Department of Physics, Harvard University, Cambridge, Massachusetts 02138, USA}

\author{M. D. Lukin}
\affiliation{Department of Physics, Harvard University, Cambridge, Massachusetts 02138, USA}

%
%
\maketitle

\section{Optical Rabi Oscillation Decoherence Measurement}
\label{eq.Rabi Oscillation Decoherence Measurement}

\begin{figure}[H]
\begin{center}
\includegraphics[width=0.7\columnwidth]{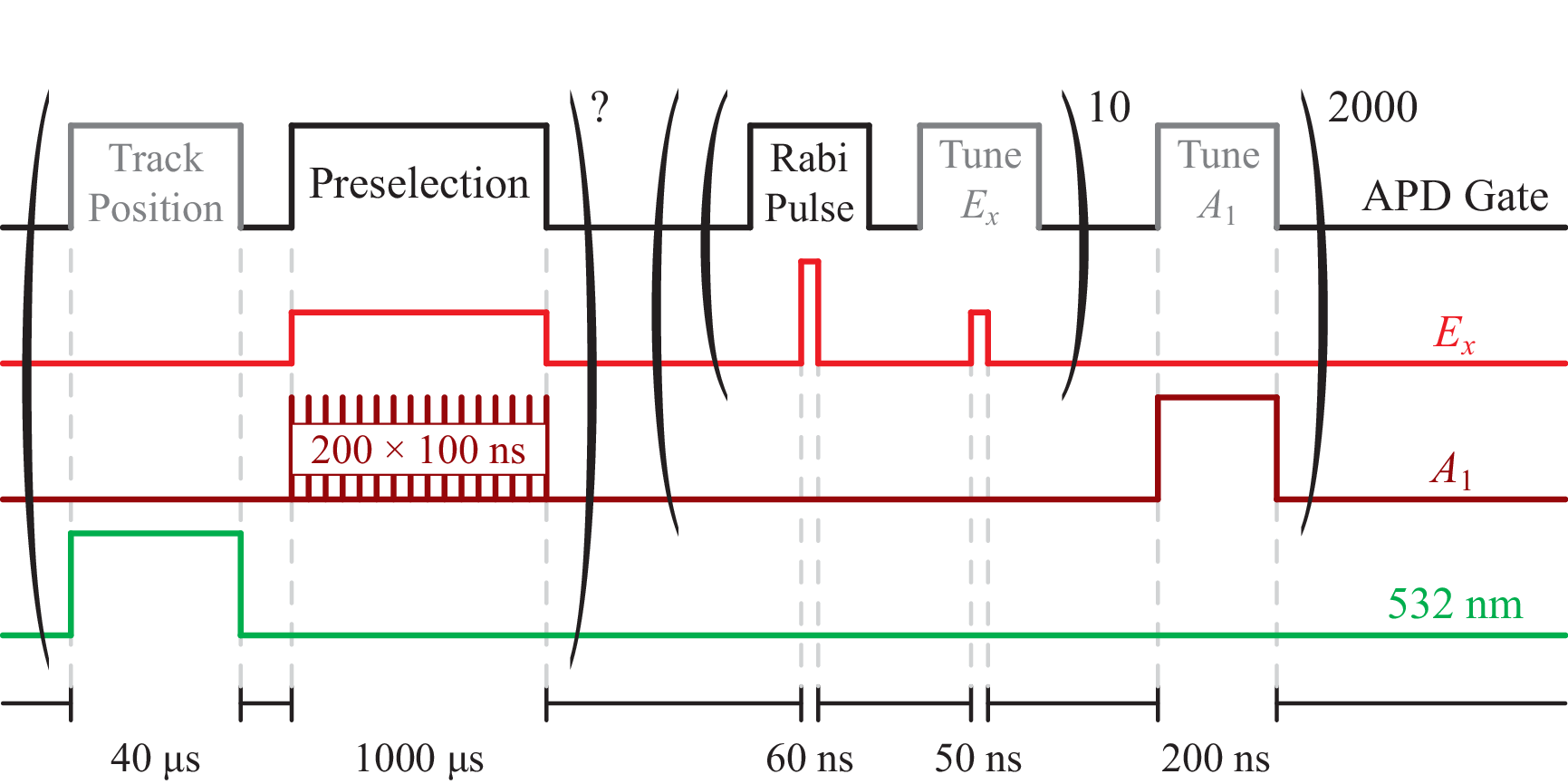}
\end{center}
\caption{The pulse sequence used to measure optical Rabi decoherence.  In the first stage, we initialize the NV center into the negatively charged state and $|0\rangle$ electronic state with nonresonant excitation at 532 nm, then we apply resonant excitation with reduced intensity/duty cycle to determine whether the NV center's transitions are on resonance.  The first stage is repeated until the number of photons collected during the preselection period surpasses a specified threshold.  In the second stage, we strongly excite the $|0\rangle-\Ex$ transition ten times before repumping on the $|\pm1\rangle-\Aone$ transition.
Every few minutes during the experiment, we compensate for slow drifts by measuring the PSB fluorescence during the periods shown in grey: we weakly excite the $|0\rangle-\Ex$ transition to tune the excitation laser precisely, we excite the $|\pm1\rangle-\Aone$ transition to tune the repump laser, and we count the photons emitted during nonresonant initialization to steer our optical path to track the NV center's position.}
\label{fig.Rabi Pulse Sequence}
\end{figure}

To measure the decoherence of optical Rabi oscillations, we apply the pulse sequence shown in Fig. \ref{fig.Rabi Pulse Sequence}.  The application of green light initializes the NV center to the negatively charged state by ionizing local charge traps in the diamond \cite{Bassett2011}, which shifts the local electric field and, through the DC Stark effect, induces spectral diffusion of the NV center's optical transitions. We negate this spectral diffusion with a preselection stage that tests whether the NV center's transitions are resonant with the excitation lasers \cite{Robledo2010}.  During the strong excitation pulse, we measure the detection times of photons in the phonon sideband (PSB) relative to the pulse beginning.  The spontaneous emission rate into the PSB is instantaneously proportional to the population in $\Ex$, enabling us to measure directly the decoherence of Rabi oscillations and decay via spontaneous emission after the end of the pulse.  We repeat this procedure at many temperatures between 5.8 K and 24 K.  Typical results for three temperatures are shown in Fig. 2(a).

As described in the main text, we extract $\GAdd$, the additional decoherence rate of the Rabi oscillations due to processes other than spontaneous emission to $|0\rangle$, from each experiment iteration.  We fit the extracted values of $\GAdd$ to
\begin{equation}
\GAdd\left(T\right)=A\left(T-T_0\right)^5+C
\end{equation}
to extract the fit constant values
\begin{align}
A &=2\pi\times\left( 2.0 \pm 0.9 \right) 10^{-5}\,\mathrm{MHz}/\mathrm{K}^5 \\
T_0 &= 4.4 \pm 1.5 \,\mathrm{K} \nonumber \\
C &= 2\pi\times\left( 0.08 \pm 0.56 \right) \,\mathrm{MHz}, \nonumber
\end{align}
where the uncertainties are the $95\%$ confidence interval bounds on the fit parameters.  The best fit and the $95\%$ confidence bands are shown in Fig. 2(a) in the main text.  We also conducted this experiment repeatedly at 5.8 K to measure $\GAdd=-0.34\pm1.87$ MHz, where the uncertainty is given by twice the standard deviation of the extracted $\GAdd$ values.

In order to extract a value for $\eta$, which parameterizes the electron-phonon coupling strength in our model of the ISC mechanism, we set $A$ equal to the $T^5$ coefficient of the $\Ex-\Ey$ mixing rate
\begin{equation}
\label{eq.Ex-Ey Mixing Rate}
\Gamma_\mathrm{Mix} = \frac{64}{\pi} \, \hbar \: \alpha \, \eta^2 k_B^5 \, T^5,
\end{equation}
where $\alpha=25.9$ is a numeric constant, that is calculated in Ref. \cite{Goldman2014a}.  We find $\eta=2\pi\times\left(44.0\pm2.4\right)\,\mathrm{MHz}\:\mathrm{meV}^{-3}$.

\section{Rabi Decoherence Analysis}
\label{sec.Rabi Decoherence Analysis}

\begin{figure}[H]
\begin{center}
\includegraphics[width=0.35\columnwidth]{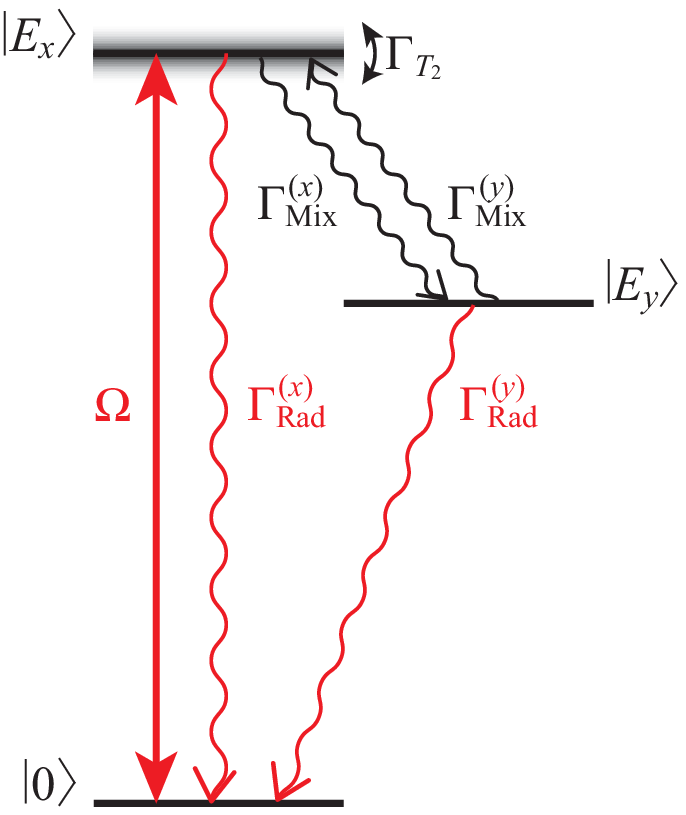}
\end{center}
\caption{The three-level system used to investigate population mixing between $\Ex$ and $\Ey$.  The directional mixing rates $\Gxy$ and $\Gyx$, radiative decay rates $\GRadx$ and $\GRady$, and decoherence rate $\GTTwo$ on the $|0\rangle-\Ex$ transition are shown.}
\label{fig.Ex-Ey State Diagram}
\end{figure}

In order to extract the phonon-induced mixing rate from the Rabi decoherence data, we must understand how the various rates factor into the Rabi decoherence timescale.  To that end, we solve the master equation in Lindblad form for the three level system shown in Fig. \ref{fig.Ex-Ey State Diagram}.  For the sake of generality, we label the two mixing rates and the two radiative decay rates separately according to the initial states of the respective processes.

We find that the emitted fluorescence ($\propto \rho_{xx}+\rho_{yy}$) oscillates within an envelope given by
\begin{equation}
\label{eq.Rabi_envelope}
g(t)=\frac{1}{2}\left(\mathrm{e}^{-t/\tRabi}+A\mathrm{e}^{-t/\tau_2}+B\right),
\end{equation}
where the exponential timescales are
\begin{equation}
\label{eq.Rabi_decay_timescale}
\frac{1}{\tRabi}=\frac{3}{4}\GRadx+\frac{1}{2}\left(\Gxy+\GTTwo\right)
\end{equation}
and
\begin{equation}
\label{eq.Rabi envelope t2}
\frac{1}{\tau_2}=\frac{1}{2}\left(2\GRady+\Gxy+2\Gyx\right),
\end{equation}
and the coefficients are
\begin{equation}
\label{eq.Rabi envelope A}
A=\frac{-\Gxy}{2\GRady+\Gxy+2\Gyx}
\end{equation}
and
\begin{equation}
\label{eq.Rabi envelope B}
B=\frac{2\left(\GRady+\Gxy+\Gyx\right)}{2\GRady+\Gxy+2\Gyx}.
\end{equation}

Because $\Ex$ and $\Ey$ are separated by $3.9\,\mathrm{GHz}\ll k_BT/2\pi\hbar$, we assume that $\Gxy=\Gyx$, which sets an upper limit of $A\leq\frac{1}{3}$.  We therefore neglect the term of $g\left(t\right)$ that decays on a timescale of $\tau_2$ and we set the measured Rabi oscillation decoherence timescale equal to $\tRabi$.  We rearrange Eq. \ref{eq.Rabi_decay_timescale} to find
\begin{equation}
\Gxy+\GTTwo=2\left(\frac{1}{\tRabi}-\frac{3}{4}\GRad\right),
\end{equation}
which is reproduced with $\Gxy=\GMix$ as Eq. 1 in the main text.

If we set $\Gxy,\Gyx\rightarrow0$, then we recover the standard result \cite{Cohen-Tannoudji1992}
\begin{equation}
g^\prime(t)=\frac{1}{2}\left(1+\mathrm{e}^{-t/\tRabi^\prime}\right)
\end{equation}
with
\begin{equation}
\frac{1}{\tRabi^\prime}=\frac{3}{4}\GRad+\frac{1}{2}\GTTwo.
\end{equation}

\section{Fluorescence Depolarization Measurement}

ZPL photons emitted by decay from the $\Ex$ and $\Ey$ states have orthogonal linear polarizations in the plane orthogonal to the N-V axis.  Therefore, we can use a polarizer in the collection path to preferentially collect fluorescence from either transition while suppressing fluorescence from the other.  This technique enables us to use depolarization of the NV center fluorescence to measure mixing of the electronic state population.

Our polarization selectivity, however, is not perfect.  Because the N-V axis lies along the [111] crystallographic axis but we collect fluorescence emitted primarily along the [100] axis, the two polarizations are not perfectly orthogonal in the lab frame.  Additionally, dichroic filters in the optical path substantially rotate polarizations that are not aligned either vertically or horizontally.  As a result, there is no perfect set of polarization settings for the experiment.  Instead, we must balance our simultaneous needs to suppress fluorescence from the undesired transition, collect fluorescence from the desired transition efficiently, suppress reflections of the strong excitation pulse, and excite the $|0\rangle\rightarrow\Ex$ transition efficiently.

We apply a pulse sequence similar to that shown in \ref{fig.Rabi Pulse Sequence}, except that we now record the arrival times of photons emitted into the ZPL instead of the PSB.  Also, we now apply a short ($\sim2$ ns FWHM) pulse to excite the NV center efficiently into the $\Ex$ state instead of applying a a long (60 ns) pulse to observe multiple Rabi oscillations.  We perform this procedure twice, with the collection optics set to collect the fluorescence from either $\Ex$ or $\Ey$.  We also repeat this procedure in both configurations with the green reionization pulse disabled in order to measure the background due to pulse reflections, ambient light, and APD dark counts. We reject photons collected before 3.3 ns after the end of the excitation pulse to further remove effects due to pulse reflections, as shown in the inset to Fig. \ref{fig.Fluorescence Depolarization}.

\section{Fluorescence Depolarization Analysis}

\begin{figure}
\begin{center}
\includegraphics[width=0.6\columnwidth]{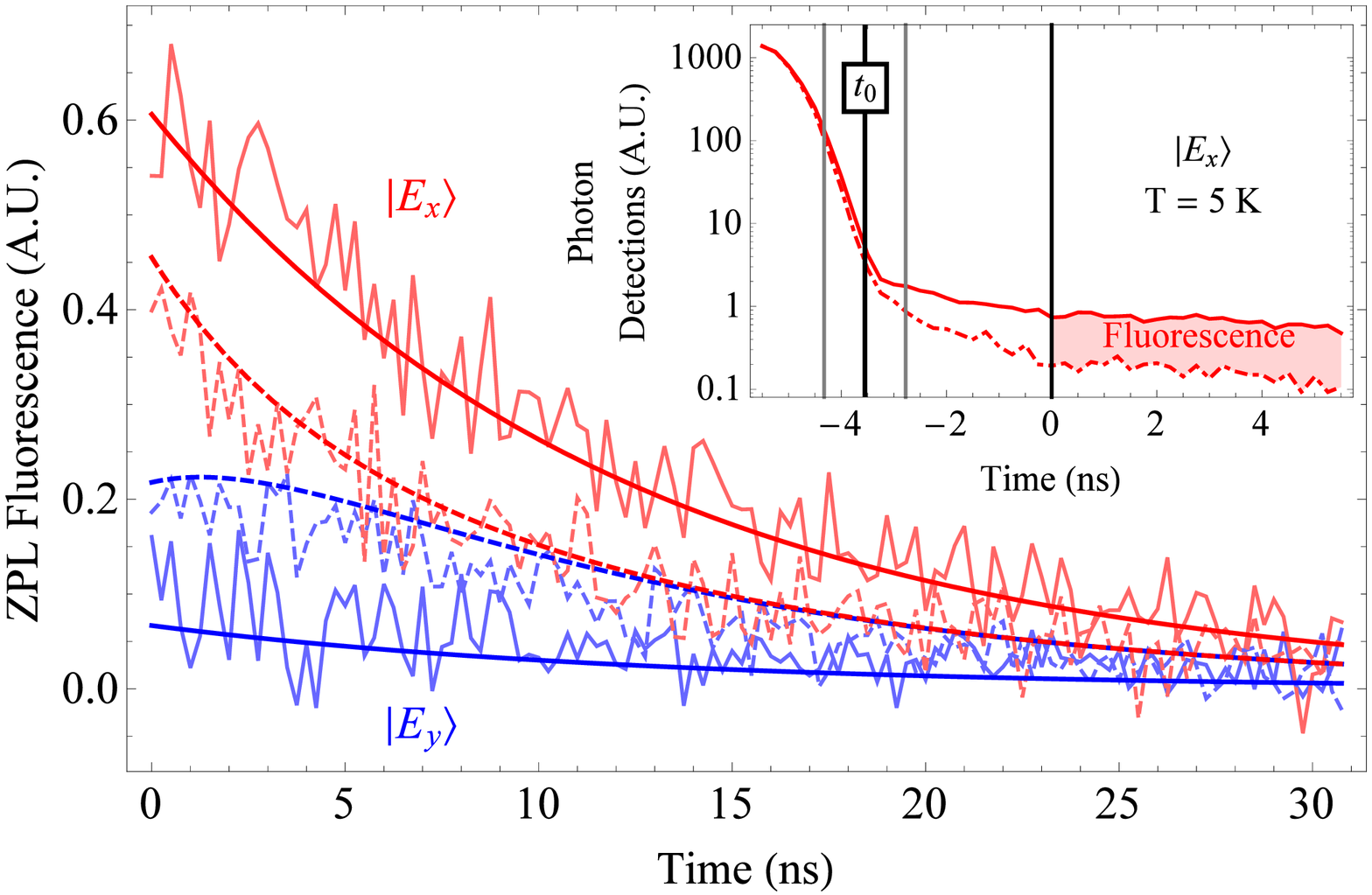}
\caption{Background-subtracted fluorescence of $x$ (red) or $y$ (blue) polarization collected after resonant excitation to $\Ex$. The data, which are also shown in Fig. 2(b) in the main text, were taken at $T=5.0$ K (dashed lines) and $T=20$ K (solid lines). The fits are simulations to the three-level system described below.  The inset, which shares a common $t=0$ ns point with the main graph, shows the $x$-polarized photons collected with (solid) and without (dashed) the green reionization pulse.  We reject all photons collected in the first 3.3 ns after the end of the reflected excitation pulse, as indicated by the shaded region beginning at $t=0$ ns.  We indicate the starting point $t_0$ of the mixing/radiative decay dynamics (with $95\%$ confidence bounds), as predicted by fitting the fluorescence data to the three-level model described in the text.}
\label{fig.Fluorescence Depolarization}
\end{center}
\end{figure}

We would like to test whether the Rabi decoherence and fluorescence polarization measurements give a consistent picture of population mixing.  We will take the values of $\GMix$ that we extract from the Rabi decoherence data and use them to simulate the fluorescence depolarization.  For the sake of simplicity, we restrict our analysis to after the excitation pulse and we ignore mixing dynamics during the pulse.  Therefore, we assume that the NV has efficiently been excited into $\Ex$ and there is initially no population in $\Ey$.

We consider a three-level system: the bright state $|B\rangle$ corresponds to $\Ex$, the dark state $|D\rangle$ corresponds to $\Ey$, and the ground state $|G\rangle$ corresponds to $|0\rangle$. We solve the population rate equations
\begin{align}
\dot{\rho}_B &= -\GRad \: \rho_B - \GMix\left(\rho_B-\rho_D\right)\\
\dot{\rho}_D &= -\GRad \: \rho_D + \GMix\left(\rho_B-\rho_D\right), \nonumber
\end{align}
where $\GRad$ is the radiative decay rate from both $|B\rangle$ and $|D\rangle$ to the ground state and $\GMix$ is the mixing rate between $|B\rangle$ and $|D\rangle$, with the initial conditions
$\rho_B\left(0\right)=1, \: \rho_D\left(0\right)=0$ to find
\begin{align}
\rho_B\left(t\right) &= \frac{1}{2}e^{-\GRad t}\left(1+e^{-2\GMix t}\right) \\
\rho_D\left(t\right) &= \frac{1}{2}e^{-\GRad t}\left(1-e^{-2\GMix t}\right). \nonumber
\end{align}

To fit the observed fluorescence, we need to account for the imperfect polarization selectivity.  We fit the fluorescence data
shown in Fig. \ref{fig.Fluorescence Depolarization} to
\begin{align}
\tilde{\rho}_B\left(t\right) &= A\left[\left(1-\epsilon\right)\rho_B\left(t-t_0\right) + \epsilon \: \rho_D\left(t-t_0\right)\right] \\
\tilde{\rho}_D\left(t\right) &= A\left[\left(1-\epsilon\right)\rho_D\left(t-t_0\right) + \epsilon \: \rho_B\left(t-t_0\right)\right], \nonumber
\end{align}
where $\epsilon$ is the error in our polarization selectivity.  We extract the mixing rates $\GRad\left(5.0\,\mathrm{K}\right)=2\pi\times0.08\,\mathrm{MHz}$ and $\GRad\left(20\,\mathrm{K}\right)=2\pi\times18.5\,\mathrm{MHz}$ from the fit to the Rabi decoherence data described in section \ref{eq.Rabi Oscillation Decoherence Measurement}.
We fit all four data sets simultaneous, using these two values of $\GRad$ and a common set of fit parameters, finding
\begin{align}
A &= 0.90\pm0.06 \nonumber \\
t_0 &= -3.6\pm0.8\,\mathrm{ns} \\
\epsilon &= 10\pm2\%. \nonumber
\end{align}

We find excellent agreement between the simulation fit and our data.  The polarization selectivity of $1-\epsilon=90\%$ is roughly consistent with our expectations of the collection path's performance.  The nonnegligible value of $\epsilon$ reflects the necessary tradeoffs inherent our choice of polarization settings, as discussed in the previous section.

The value of $t_0$ is also consistent with our expectations.  Our simplified model of an undriven three-level system subject only to radiative decay and nonradiative population transfer is necessarily valid only after the end of the excitation pulse.  The value of $t_0$ extracted from the fit places $t_0$ near the end of the excitation pulse's falling edge, as shown in the inset to Fig. \ref{fig.Fluorescence Depolarization}.  Essentially, the simulation, when extrapolated backward toward the excitation pulse, picks out the time that marks the beginning of the underlying model's validity.  A more precise statement would require a model that incorporates the resonant driving dynamics that occur during the excitation pulse as well as the nonnegligible ($\sim2$ ns) width of the pulse's falling edge, which is beyond the scope of this analysis.  The success of the simplified model, however, provides strong evidence that the Rabi decoherence and fluorescence polarization measurements give a consistent picture of population mixing.

\section{Excited State Lifetime Measurement}

To measure the lifetimes of $|A_1\rangle$, $|A_2\rangle$, and $|E_{1,2}\rangle$, we again employ an experimental procedure similar to that depicted in Fig. \ref{fig.Rabi Pulse Sequence}.  In this case, however, the primary excitation laser, labeled ``$E_x$'' in the figure, is tuned to the transition between $\pm1$ and one of the $\Et$ states listed.  We apply a short excitation pulse, as in the fluorescence depolarization measurement.  We repump on the $|0\rangle\rightarrow\Ey$ transition for 10 $\mathrm{\mu s}$ to repopulate the $|\pm1\rangle$ states; we repump for a longer time because the $\Ey$ transition is more closed than the $\Aone$ transition previously used to repump to the correct spin state.  We perform this procedure at several temperatures between 5 K and 26 K.

To extract the excited state lifetimes, we fit each dataset to a single-exponential decay function.  The fit window starts 4 ns after the beginning of the excitation pulse and extends for 115 ns.  This 4 ns delay was selected to remove any effects of the excitation pulse, ensuring that the fluorescence we consider is solely the result of spontaneous emission.

\section{ISC Rate from $\Ex$}
\label{sec.Neglecting GISCx}

We now justify the assumption that the ISC rate from $\Ex$ is negligible, which enables the state-dependent ISC rates shown in Fig. 3 in the main text to be extracted from the measured fluorescence lifetimes.  The ratio of the ISC rate from the $\Et$ states with $|m_s|=1$ ($\GISCone$) to the ISC rate from the $\Et$ states with $m_s=0$ ($\GISCzero$) has been addressed both theoretically and experimentally \cite{Manson2006,Jelezko2004,Robledo2011}.

Manson \textit{et al.} \cite{Manson2006} developed a detailed model of NV center spin dynamics from a careful consideration of the NV center's symmetry properties.  Their model predicts $\GISCzero=0$, a conclusion that is consistent both with their measurements of the NV center's transient behavior under nonresonant optical excitation and with the conclusions of an earlier review of nonresonant spin initialization and readout \cite{Jelezko2004}, which cited $\GISCzero\sim10^3 \,\mathrm{s}^{-1}$ and $\GISCone\sim10^6 \,\mathrm{s}^{-1}$.  This conclusion is somewhat inconsistent, however, with the more recent observations of Robledo \textit{et al.} \cite{Robledo2011}, who used nonresonant excitation to measure the spin-dependent lifetimes of the $\Et$ states, the lifetime of the metastable $\Es$ states, and the degree of spin polarization in the $\Et$ manifold.  They used these measurements to construct a phenomenological model of NV center dynamics, from which they extracted $\GISCzero/2\pi\approx1-2$ MHz.  Similarly, Tetienne \textit{et al.} \cite{Tetienne2012} applied measurements of photoluminescence intensity, ESR contrast, and $\Et$ state lifetimes as functions of an off-axis magnetic field to the same model to extract $\GISCzero/2\pi\approx0.8-1.7$ MHz.

The question of $\GISCx$ is, in general, complicated by the facts that there is significant phonon-induced depolarization between $\Ex$ and $\Ey$ for $T>15$ K and that the $\Ey$ and $\Eonetwo$ states exhibit a level anticrossing when the strain-induced $\Ex-\Ey$ splitting is approximately 7 GHz \cite{Robledo2011a}.  Thus, phonons couple $\Ex$ to $\Ey$, spin-spin interaction couples $\Ey$ to $\Eonetwo$, and $\Eonetwo$ decay to $\As$ through the ISC mechanism described in Ref. \cite{Goldman2014a}.  The $\GISCx$ due to such a mechanism would depend on both temperature and crystal strain.  Because both Robledo and Tetienne considered the NV center at $T=300$ K in their models and neither specified the $\Ex-\Ey$ splitting, this mechanism could be responsible for their non-negligible values of $\GISCzero$.  In the context of this work, however, the fact that the measured lifetime of $\Ex$ does not depend on temperature from 5 K, where mixing between $\Ex$ and $\Ey$ is negligible, to 26 K, where mixing is much faster than radiative decay, indicates that the contribution of this mechanism is negligible.

\begin{figure}[h]
\begin{center}
\includegraphics[width=0.35\columnwidth]{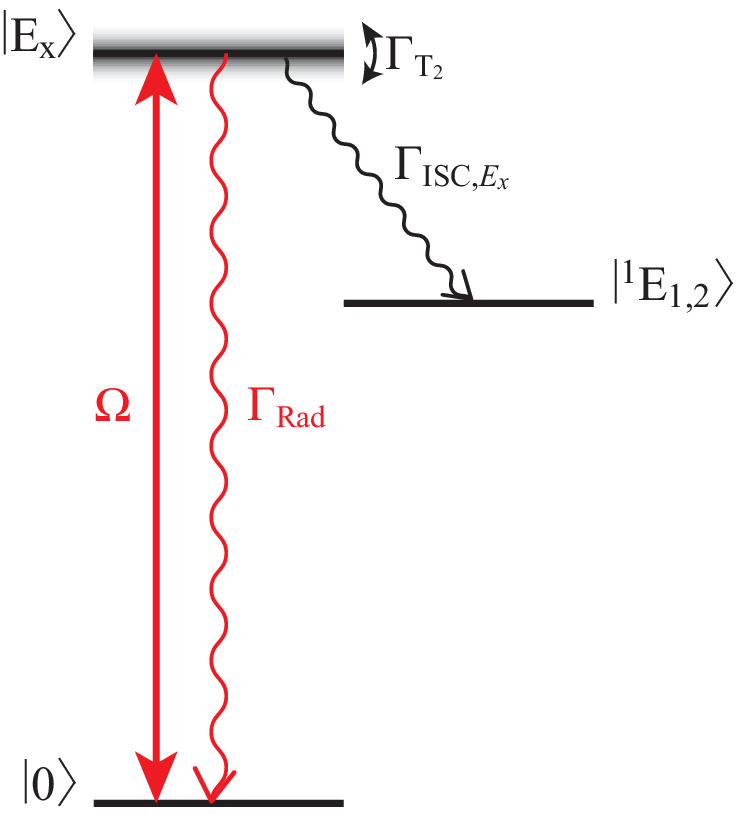}
\end{center}
\caption{The three-level system used to extract $\GISCx$ from a measurement of optical Rabi oscillations conducted at 5.8 K.  Because the oscillation and ISC dynamics that occur over a timescale of $\sim4$ to 60 ns and the lifetimes of the singlet states are $\tau_{^1A_1}=0.9$ ns \cite{Acosta2010a} and $\tau_{^1E_{1,2}}\sim370$ ns \cite{Robledo2011}, we assume that population neither returns from $\As$ to $\Ex$ nor decays from $\Es$ to $|0\rangle$ once it has undergone the ISC transition to the singlet states.  Thus, we can combine $\As$ and $\Es$ into one effective dark state.}
\label{fig.Ex ISC States}
\end{figure}

We can also place a limit of $\GISCx$ based on the measured coherence time of optical Rabi oscillations.  If we consider the low-temperature limit, where mixing between $\Ex$ and $\Ey$ is suppressed, then we can perform an analysis analogous to that described in Sec. \ref{sec.Rabi Decoherence Analysis}, where we substitute the ISC crossing to the singlet states for phonon-induced mixing to $\Ey$.  The resulting three-level system is shown in Fig. \ref{fig.Ex ISC States}.  We solve the corresponding master equation in Lindblad form and find that the emitted fluorescence ($\propto \rho_{E_xE_x}$) oscillates within an envelope given by
\begin{equation}
\label{eq.Rabi envelope with ISC}
g(t)=\frac{1}{2}\left(e^{-t/\tRabi}+1\right)e^{-\GISCx t/2},
\end{equation}
where the exponential timescale is
\begin{equation}
\label{eq.tRabi with ISC}
\frac{1}{\tRabi}=\frac{3}{4}\GRadx+\frac{1}{2}\GTTwo.
\end{equation}

We fit the fluorescence during the Rabi oscillations to
\begin{equation}
f\left(t\right) = A\left[\cos\left(\Omega \, t - \phi\right)e^{-(t-t_0)/\tRabi}+1\right] e^{-\GISCx t/2},
\end{equation}
where all quantities except for $t$ are free fit parameters.  The extracted initial time $t_0$ of the Rabi oscillation decay is found to be close (within $3 \,\mathrm{ns}\sim\tau_\pi$) to the start of the excitation pulse in all cases, indicating that the visibility of the Rabi oscillations is well described by Eq. \ref{eq.Rabi envelope with ISC}.

From these fits, we extract $\GISCx/2\pi=0.62\pm0.21$ MHz.  We note that our model does not take into account spin non-preserving radiative decay from $\Ex$ into the dark $|\pm1\rangle$ $\At$ states or deionization into the dark NV$^0$ charge state, both of which would mimic the effect of the ISC transition into the dark metastable singlet state.  Thus, this value represents an upper bound on $\GISCx$ that is an order of magnitude lower than the measured $\GISCone$.  We can therefore assume that $\GISCx$ is negligible, in agreement with the preponderance of literature cited above.

\section{ISC Rate Analysis}

In Fig. 3 in the main text, we fit the ISC rates observed after excitation into $\Aone$ or $\Atwo$ to a simple model of phonon-induced state mixing.  In this model, we assume the following sequence of events.  At $t=-t_0$, we instantaneously transfer the entire population to the target state with a perfect $\pi$ pulse.  Phonons induce state mixing at a rate $\Gamma_\mathrm{Mix}$ until $t=0$, at which point we begin fitting the fluorescence decay.  Mixing continues at a rate $\Gamma_\mathrm{Mix}$ throughout the entire fitting period, from $t=0$ to $t=\Delta t$.  We then fit the PSB fluorescence observed during the measurement period to a simple exponential.  Values of $t_0=4$ ns and $\Delta t=115$ ns were chosen to match the analysis performed on the experimental data, as described in the previous section.

To model the dynamics of the system under mixing between $\Aone$ and $\Atwo$, radiative decay from both states, and ISC decay from $\Aone$ alone, we solve
\begin{align}
\dot{\rho}_{A_1} &= -\left(\GRad + \GISC\right) \rho_{A_1} - \GMix\left(\rho_{A_1}-\rho_{A_2}\right) \nonumber \\
\dot{\rho}_{A_2} &= -\GRad \: \rho_{A_2} + \GMix\left(\rho_{A_1}-\rho_{A_2}\right).
\end{align}
The quantity of interest is the measured fluorescence intensity, which is proportional to $\rho_{A_1}+\rho_{A_2}$.  The fluorescence intensity measured after excitation into $A_{1,2}$ is given by
\begin{equation}
\label{eq.Fluor I}
I_{A_{1,2}} = e^{-\left(\GRad+\GMix+\GISC/2\right)\,t}\left[\frac{2\GMix\mp\GISC}{\Gamma^\prime}
\sinh\left(\frac{\Gamma^\prime}{2}\,t\right)+\cosh\left(\frac{\Gamma^\prime}{2}\,t\right)\right],
\end{equation}
where $\Gamma^\prime = \sqrt{\GISC^2+4\GMix^2}$.

We simulate the fluorescence intensity observed during the measurement period using Eq. \ref{eq.Fluor I}.  We expect that the mixing rate between $\Aone$ and $\Atwo$ should be equal to that between $\Ex$ and $\Ey$ \cite{Goldman2014a}.  We therefore set the temperature-dependent $\GMix$ equal to the value given by our fit to the $\Ex-\Ey$ mixing data, which is described in Sec. \ref{eq.Rabi Oscillation Decoherence Measurement}.

We then apply the same analysis that we used to extract $\GISC$ from our measured fluorescence intensity data, giving the effective ISC rates measured after excitation into $\Aone$ or $\Atwo$ as functions of temperature and $\GAone$.  We perform a $\chi^2$ minimization fit to the measured temperature-dependent ISC rates to find $\GAone/2\pi = 16.0\pm0.6$ MHz.


%